\documentclass[sigconf, 10pt]{acmart}

\usepackage{stfloats}
\usepackage{url}
\usepackage{verbatim}
\usepackage[utf8]{inputenc}
\usepackage{graphicx}
\usepackage{amsmath}
\usepackage{amsfonts}

\usepackage{amssymb}
\usepackage{algpseudocode}
\usepackage{algorithm}
\graphicspath{{images/}{../images/}}
\usepackage{subfiles}
\usepackage{fancyhdr}
\usepackage{xcolor}
\usepackage{multirow}
\usepackage{textcomp}
\usepackage{array}
\usepackage{tabu}
\usepackage{amsmath}
\usepackage{bm}
\usepackage{caption}
\usepackage{subcaption}
\usepackage{enumitem}
\usepackage{booktabs}       
\usepackage{pifont}         
\usepackage{makecell}       

\def \method {Eunoia}
\def \rma {RAN Monitor}
\def \spa {SM Predictor}
\def \saa {SM Allocator}

\usepackage{graphicx,color}
\usepackage{hyperref}

\usepackage{colortbl}      
\usepackage{fontawesome5}  
\usepackage{tabularx}

\setcopyright{rightsretained}

\acmDOI{10.475/123_4}

\acmISBN{123-4567-24-567/08/06}

\acmConference[WOODSTOCK'97]{ACM Woodstock conference}{July 1997}{El
  Paso, Texas USA} 
\acmYear{1997}
\copyrightyear{2016}

\acmArticle{4}
\acmPrice{15.00}

\editor{Jennifer B. Sartor}
\editor{Theo D'Hondt}
\editor{Wolfgang De Meuter}

\renewcommand\footnotetextcopyrightpermission[1]{}

\begin{document}
\fancyhead[LE,RO]{}
\title{\method: Toward Application-Transparent, Scalable GPU Sharing in AI-RAN}


\author{Zongshen Wu, Marko Milenković, Jiarong Xing, Rahman Doost-Mohammady}
\affiliation{%
  \institution{Rice University}
  \city{Houston} 
  \state{TX}
  \country{US}
}







\renewcommand{\shortauthors}{ }

\begin{abstract}
The AI-RAN paradigm makes the next-generation cellular network more versatile by co-locating AI and RAN workloads on the same commodity hardware. While \textcolor{black}{the} GPU is a promising platform for AI-RAN realization, a generic concurrent AI application interferes with the latency-sensitive RAN on GPU compute \textcolor{black}{resources}. In this paper, we propose \method, a dynamic real-time AI-RAN GPU scheduler that is transparent to both AI applications and the RAN workload and operates at the granularity level of the RAN slot. We deploy \method\ onto NVIDIA Aerial AI-RAN and evaluate \method\ on both an over-the-air 5G private network and an emulated multi-cell network. We demonstrate that in a 27-cell AI-RAN, \method\ keeps the RAN's uplink $P_{99}$ latency under 3 ms while delivering higher AI performance than other RAN-safe static partitioning methods. 
\end{abstract}

%
%



\maketitle





\begin{figure}
    \centering
    \includegraphics[width=0.8\linewidth]{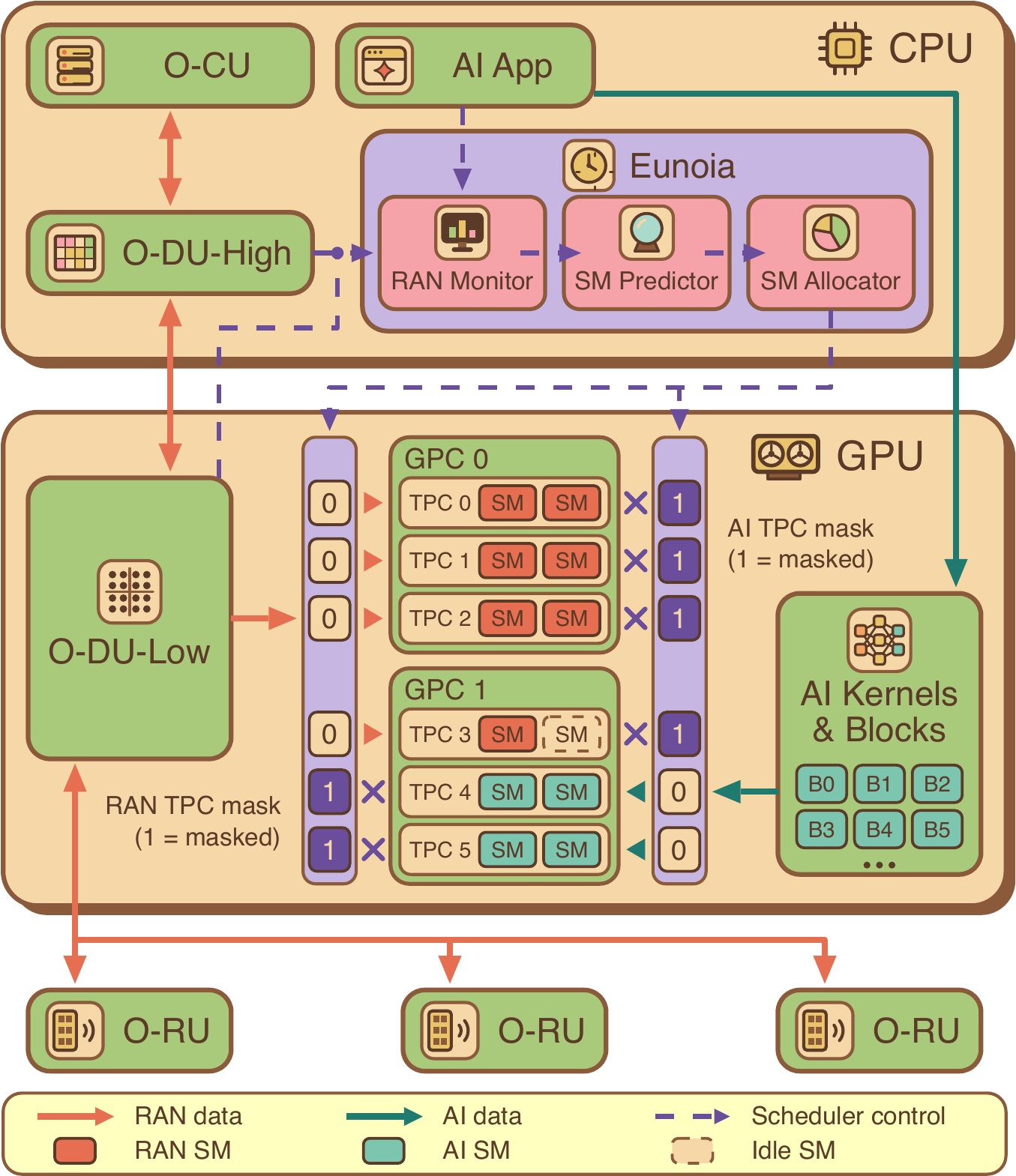}
    \caption{\method\ system architecture}
    \label{fig:sysarch}
\end{figure}

\section{Introduction}

The convergence of cellular networks and artificial intelligence (AI) is a defining force behind 6G, where intelligence moves into the last mile of connectivity, within low-latency reach of wireless users and enabling novel verticals. This paradigm, coined AI-RAN~\cite{nv-airan}, equips the radio access network (RAN) with the ability to host edge AI models, including machine learning models and large foundation models, alongside the RAN itself. Three use cases are outlined by the AI-RAN alliance~\cite{airanalliance}: (1) AI \emph{for} the RAN, e.g., real-time control for scheduling, traffic steering, and sub-millisecond inference loops inside the distributed unit (DU)~\cite{edgeric, dapp}; (2) AI \emph{on} the RAN that uses the same infrastructure to serve large language models (LLMs), video analytics, and distributed Physical AI workloads to the RAN users~\cite{arma}; (3) AI \emph{and} RAN that runs non-RAN related AI models alongside RAN. 

The use of data-center-grade GPUs for high-performance RAN baseband processing, realized by NVIDIA Aerial~\cite{aerial, aerial_sdk}, makes the AI-RAN vision practical. Aerial is capable of running the complete 5G physical layer, the O-DU-Low, for tens of 5G cells on a single Hopper or Blackwell GPU and can achieve gigabit-per-second network throughput on par with other advanced commercial cellular products. Recent studies have characterized the performance and efficiency of this approach~\cite{ethos-wintech, decodex}. Naturally, the same GPU can be used for running AI model inference, not only placing inference just one hop away from the RAN users, but also considerably improving the economics of AI-RAN. However, co-locating these workloads on the same GPU is not trivial. The RAN must deliver carrier-grade availability with per-slot processing deadlines at the DU. Uplink decoding and downlink encoding for every slot must finish within a few milliseconds; otherwise HARQ retransmissions accumulate, the user equipment (UEs) lose synchronization, and cells drop. GPU resource starvation at that time scale, even briefly, is unacceptable to an operator.

The current widely used solution for co-locating sensitive GPU workloads is through hardware-supported static partitioning. NVIDIA offers Multi-Instance GPU (MIG)~\cite{mig} that statically partitions the GPU into fixed slices at the hardware level based on a set of predefined configurations. 
Thus, each partition has to be provisioned for the worst case, which leads to underutilized resources in lower load scenarios. Multi-Process Service (MPS)~\cite{mps} is a finer-grained solution, but only enforces a per-process upper bound on streaming multi-processor (SM) usage, and does not carry any notion of priority or active provisioning~\cite{mps_pitfall}. 
Both MIG and MPS solutions force the operator to choose between RAN reliability and AI throughput, and the choice must be re-made for every change in cell count, bandwidth, or AI model mix. \textcolor{black}{The Green Context is} another recent feature introduced by NVIDIA that allow SM partitioning with sub-process granularity. However, they require modifications to each application code, which makes them difficult to use in a heterogeneous AI-RAN application environment.

Compute sharing for virtualized RAN (vRAN) has been explored on CPUs~\cite{concordia, nuberu}. CPU vRAN sharing relies on core pinning and OS scheduling primitives which do not directly translate to GPUs. None of them addresses the AI-RAN setting directly. YinYangRAN~\cite{yinyangran} predicts the traffic patterns and adjusts the MPS management through orchestration for AI-RAN. 
However, reconfiguring MPS settings is not real-time (>100 ms) and requires workload fallback on CPU during the reconfiguration.
Recent work has studied fine-grained GPU sharing among AI workloads, given the large inefficiencies in AI data centers ~\cite{lithos, orion}. However, they operate at the granularity of kernels or inference requests, target latencies of tens of milliseconds, and treat all tenants as peers. AI-RAN is posing a more challenging problem that existing methods, where the resource demand for RAN workloads changes at the slot-level (sub-millisecond) granularity and if not properly handled, it will cause RAN crashes or significant performance degradation. Therefore, there is a need for a GPU sharing framework for AI-RAN that can track RAN GPU resource demand at sub-millisecond granularity, while efficiently allocating every spare SM to the AI workloads. Weaver~\cite{weaver} proposes a framework for GPU sharing between RAN and Foundation model training. However, it uses Green Context as GPU sharing mechanism, only considers RAN LDPC decoding (look-aside RAN acceleration) on the GPU and does not generalize to many concurrent cells. 

In this paper, we present our vision for fine-grained, appli\-cation-transparent GPU sharing that scales to diverse AI-RAN workloads. Because AI-RAN can host a wide range of AI applications, it requires a versatile execution framework that can co-locate diverse AI models with equally diverse RAN workloads, such as distributed units (DUs) serving cells with different configurations. We argue that such a framework must satisfy three requirements. (1) \emph{Transparency:} it should not require modifications to RAN or AI application code, and should instead infer GPU resource demands and priorities through existing or lightweight interfaces. (2) \emph{Fine granularity in time and space:} it should infer SM demand at sub-millisecond timescales, matching RAN slot durations, and allocate resources at the granularity of a few SMs. (3) \emph{Scalability:} current data-center GPUs can run the entire PHY for tens of cells, each with a different configuration (e.g., number of antennas or bandwidth), while concurrently serving several AI models through spatial or temporal sharing. The framework must therefore scale to such deployments while ensuring that RAN workloads consistently meet their processing deadlines.

As a first step towards this vision, we design and prototype \method{}, which addresses the RAN side of these requirements. \method{} is transparent to the RAN: it reads real-time RAN metrics through the RIC and uses them to infer the DU’s SM demand without modifying RAN code. It reallocates SMs at sub-millisecond timescales and at a single TPC granularity, and it scales across many cells and varying traffic loads. Our key insight is that the GPU demand of the RAN is a function of RAN state: the number of active cells and UEs, per-cell uplink and downlink throughput, modulation and coding scheme, physical resource block allocation, and MIMO layers. All of these are observable at slot-level granularity from the O-DU through standard interfaces such as FAPI~\cite{fapi_222}, and they are known before the corresponding GPU kernels launch. A design block diagram of \method{} is illustrated in Fig.~\ref{fig:sysarch}
We evaluate \method{} with TimelyLLM~\cite{timelyllm} as a co-located AI workload. Extending transparency to the AI side remains an open challenge: when multiple AI models share the GPU, with varying levels of priority with respect to the RAN, the framework must decide how to prioritize among them and through what interface those models should express their requirements. We discuss these open questions in \S\ref{sec:conc}.

\section{Background and Motivation}

\subsection{Background}
\textbf{AI-RAN.} AI-RAN paradigm initiates the leverage
on AI applications for RANs and vice versa. Pioneer work such as Agora~\cite{agora}, Nuberu~\cite{nuberu}, and Megastation~\cite{megastation} provide approaches to support RAN on commodity hardware. Recent work like Concordia~\cite{concordia}, EdgeRIC~\cite{edgeric}, ETHOS~\cite{ethos-wintech}, and TimelyLLM~\cite{timelyllm} shed light on the AI-RAN framework design with paragon AI applications.

\noindent
\textbf{GPU Architecture.}
NVIDIA GPUs consist of different engines, with the major one being the Compute/Graphics Engine. The Compute/Graphics Engine is then subdivided into General Processing Clusters (GPCs). A GPC is comprised of multiple \textcolor{black}{Texture} Processing Clusters (TPCs), which contain two SMs each. A block is scheduled to one SM for execution, and one kernel from the AI application consists of multiple blocks. 

\noindent
\textbf{GPU resource partition \& scheduling.}
NVIDIA provides its official GPU resource partition tools, including MIG~\cite{mig} and MPS~\cite{mps}.
Meanwhile, other GPU resource partition and scheduling methods have emerged, typically for AI workloads like Deep Neural Network (DNN) and LLM. 
libsmctrl~\cite{libsmctrl, mps_pitfall} proposes a light-weight disaggregated run-time SM partition method for CUDA process and stream.
Orion~\cite{orion} arranges the order of DNN kernels regarding their offline trained GPU usage pattern for scheduling. 
Reef~\cite{reef} uses preemption to schedule the DNN workloads.
LithOS~\cite{lithos} atomizes the kernels to alleviate head-of-line blocking for TPC scheduling with resizing.

\definecolor{tblGreen}{HTML}{1E8E3E}
\definecolor{tblRed}{HTML}{D93025}
\definecolor{tblAmber}{HTML}{E37400}
\newcommand{\cmark}{\textcolor{tblGreen}{\ding{51}}}
\newcommand{\xmark}{\textcolor{tblRed}{\ding{55}}}
\newcommand{\pmark}{\textcolor{tblAmber}{\raisebox{0.15ex}{\scalebox{0.7}{\ding{108}}}}}
 

\begin{table}[!h]
    \caption{AI-RAN GPU Scheduler Comparison}
    \centering
    \footnotesize
    \setlength{\tabcolsep}{2pt}
    \begin{tabular*}{\columnwidth}{@{\extracolsep{\fill}}lccccc@{}}
    \toprule
    \scriptsize\textbf{Mechanism} & \scriptsize\textbf{Predictive} &
    \scriptsize\textbf{\shortstack{Trans-\\parent}} & \scriptsize\textbf{\shortstack{H.W.\\granularity}} & \scriptsize\textbf{\shortstack{Real-\\time}} & \scriptsize\textbf{\shortstack{Full-PHY\\ \& Multi-cell}} \\
    \midrule
    CAORA (MIG)~\cite{caora} & \pmark & \cmark & \textcolor{tblRed}{GPC} & \xmark & \xmark \\
    YinYangRAN (MPS)~\cite{yinyangran} & \pmark & \pmark & \textcolor{tblAmber}{SM \%} & \xmark & \xmark \\
    Weaver (Green Ctx.)~\cite{weaver} & \xmark & \xmark & \textcolor{tblAmber}{8 SMs} & \cmark & \xmark \\
    \midrule
    \textbf{\method~(ours)} & \cmark & \cmark & \textcolor{tblGreen}{2 SMs} & \cmark & \cmark \\
    \bottomrule
    \end{tabular*}
    \label{tab:bg_comp}
    \vspace{-5mm}
\end{table}

\begin{figure}[!h]
    \centering
        \includegraphics[width=0.99\linewidth]{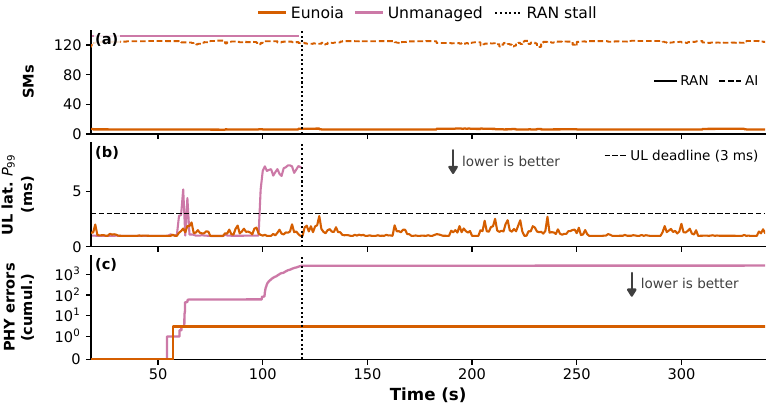}
    \caption{\textcolor{black}{Over-the-air (OTA) AI-RAN running 4-UE traffic and TimelyLLM AI application: (a) SM scheduling comparison, (b) UL process $P_{99}$ latency, (c) PHY layer cumulative errors}}
    \label{fig:eval_fullstack}
\end{figure}

\subsection{Challenge: Avoiding resource interference between AI workloads and RAN}
\label{sec:chanl}

We conduct experiments with an end-to-end 5G network set up in our lab where the NVIDIA Aerial RAN stack is running on a GH200 server with a Foxconn O-RU and an Amarisoft UE emulator. Through the UE emulator, we connect 4 UEs to the network, while simultaneously running TimelyLLM, a low-latency LLM serving system for edge robotic control. Fig.~\ref{fig:eval_fullstack} shows the various metrics for each process when run 1) without any constraint (unmanaged), and 2) with \method{} as the GPU management framework. In the unmanaged case, the RAN simply stalls at ~120 seconds while using \method{}, the RAN stably runs with the AI workload. TimelyLLM occupies the majority of SMs (Fig.~\ref{fig:eval_fullstack}(a)) and thus in the unmanaged case the uplink processing latency of the RAN (Fig.~\ref{fig:eval_fullstack}(b)) crosses the 3ms deadline and the number of PHY errors significantly increases (Fig.~\ref{fig:eval_fullstack}(c)), which causes the RAN to crash. \method{} avoids this by carefully monitoring the RAN parameters and inferring the SM usage and carefully allocating them to the RAN while allowing the rest of the SMs to be used by TimelyLLM.

\section{\method\ System Design}

In this section, we present \method\ system design, aligning with the three requirements for GPU sharing,\textit{ fine granularity, scalability, and transparency}. \method\ consists of three main modules, shown in Fig.~\ref{fig:sysarch}, \rma, \spa, and \saa.



\subsection{\method\ \rma}


We use an extension of a shared memory-based interface developed in~\cite{ethos-wintech} to develop the RAN monitor module. Through this interface, we develop the \rma~module with the O-DU as a dApp~\cite{dapp} supporting slot-level profiling granularity with sub-millisecond overhead. We measure a variety of RAN metrics, including cell-level parameters such as number of active cells and their respective UEs, and per-user metrics (e.g., MCS, PRB, throughput, TBS). The O-DU-Low running on the \textcolor{black}{GPU} supports multiple RAN cells within one process which allows monitoring all cells simultaneously through the same interface.  
Separately, the \rma~module also profiles the GPU compute resource usage using Nsight~\cite{nvidia2026nsys} with sub-millisecond slot level granularity. Since Nsight is an offline profiling tool, we will use it for fine-granularity SM usage profiling and offline training incorporating both wireless and GPU usage metrics.


\subsection{\method\ \spa}

The predictor is required to support versatile multi-cell cases with scalability, which is challenging since the metrics dimension varies along the cell numbers. Therefore, \spa\ aggregates the input over cells, in order to serve different cell counts, UE loads, and TDD patterns without retraining.
The \spa~module predicts the RAN's per-slot SM demand from the metrics exposed by the \rma~module and passes it to the \saa~module. It is an ensemble of four multi-layer perceptrons (3 hidden layers of 128 ReLU units each), whose predictions are averaged. Each slot's inputs are normalized cell-aggregated DL and UL load metrics: PRB×layers, transport-block bytes and count, the number of active cells, and for UL also the MCS-weighted load and the largest per-cell load. They are taken over a short window of slots around the target slot, together with their 20-slot running means, the TDD slot position, and the cell configuration (cell count and DMRS symbols). The networks are trained to predict the 99.8th percentile of the next slot SM usage instead of the mean as a conservative measure for RAN safety. The training labels are per-slot SM usage measured with Nsight Systems on emulated, but fully realistic, workloads. At run time, the predictor needs only the wireless metrics. Inference is implemented in C on the CPU, which provides minimal overhead, and the modules exchange data through the dApp interface.


\subsection{\method\ \saa}
The \saa\ should be transparent to both the RAN stack and AI applications for practicality. 
The \saa~ converts the SM prediction into TPC mask indexes, and then applies these indexes accessibility through libsmctrl~\cite{libsmctrl}. 
The TPC mask limits the accessible GPU SMs of a kernel launch at run time.
The \saa\ controls the GPU SM accessibility for both the RAN process and the AI application, through their individual TPC masks. Therefore, the RAN process and the AI application utilize their own isolated shares on the GPU SM resource, avoiding SM contentions.
In this case, the \saa~module dynamically manages the hardware SM partition between the latency-sensitive RAN process and the AI application, with per-slot time granularity and per-TPC GPU spatial granularity. To the best of our knowledge, \method\ is the first to implement such \textcolor{black}{fine-grained} GPU SM scheduling for AI-RAN.

\begin{figure}[!h]
        \centering
        \includegraphics[width=0.99\columnwidth]{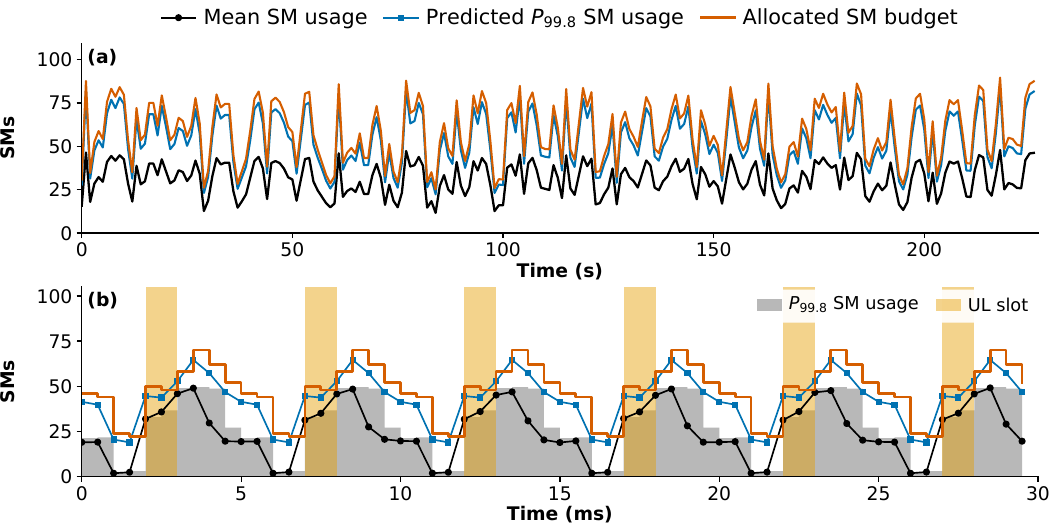}
    \caption{\textcolor{black}{\method\ prediction and allocation with the 27-cell case: (a) overview level, (b) per-slot level}}
    \label{fig:eval_ml}
\end{figure}

\begin{figure}[!h]
    \centering
        \includegraphics[width=0.99\linewidth]{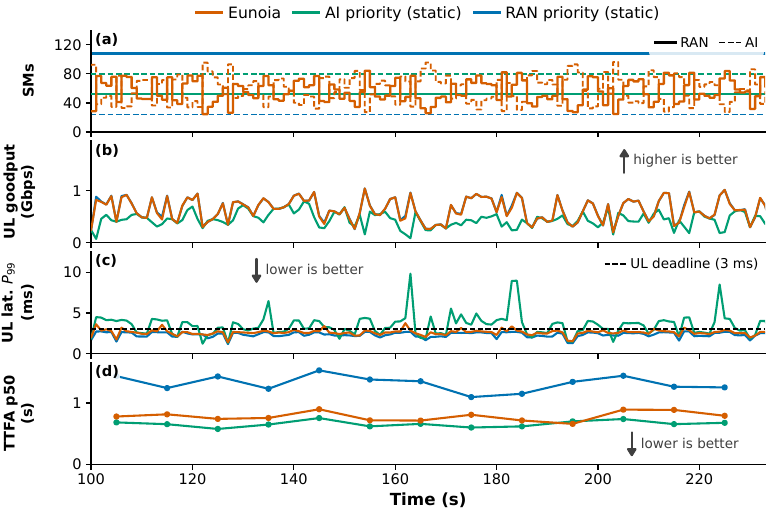}
    \caption{\textcolor{black}{Scaled-up multi-cell AI-RAN running 27 cells and TimelyLLM AI Application: (a) SM scheduling comparison, (b) RAN UL goodput, (c) UL process $P_{99}$ latency, (d) TimelyLLM service TTFA}}
    \label{fig:eval_multicell}
\end{figure}

\section{Evaluation}


We first train \method\ with the training dataset collected using NVIDIA TestMAC and RU emulator, which are parts of NVIDIA ARC-OTA 1.7 platform~\cite{aerial}. The TestMAC and RU emulator run on two NVIDIA GH200 servers connected together, which support emulating dynamic scenarios with multiple cells as well as scaled-up UE numbers and traffic load.
Then, we deploy the trained \method\ onto an over-the-air (OTA) private 5G network for the evaluation in a real OTA scene. Note that, we train \method\ purely based on TestMAC, and the OTA 5G network is completely unseen during the training process.
The OTA network setup uses a Foxconn 4$\times$4 MIMO O-RU with 100 MHz channel bandwidth and O-RAN 7.2a split interface, connected with a NVIDIA GH200 server. The server also runs ARC-OTA 1.7 as the RAN. We use Amarisoft SimBox UE emulator as the OTA UEs.
Moreover, we further deploy the same trained \method\ onto TestMAC for the scaled-up multi-cell scenario scheduling evaluation. Specifically, the scaled-up scenarios are also completely unseen during the training.
We apply TimelyLLM~\cite{timelyllm} as the AI application on the AI-RAN for both the OTA and multi-cell network scenarios. TimelyLLM provides low-latency LLM service for mobile robot controlling, including robot arm, quadruped, and drone, which is an AI-RAN application paradigm.



\subsection{Training and Dynamic Scheduling}
\label{sec:eval_train}
We train the MLP model in the \spa~module with the collected training dataset. Particularly, we train the model once with the training set, and then apply the trained weights to all the evaluation cases, since the model is generalized by design. \textcolor{black}{Fig.~\ref{fig:eval_ml} presents the \textcolor{black}{prediction and allocation} of \method~dynamic scheduling.
The overview figure illustrates the dynamic SM scheduling with a 220 second test case. The scenario in the test case runs up to 27 concurrent cells with fluctuated traffic patterns and UE numbers. The black line shows the actual RAN SM usage measured by Nsight. The blue and orange lines present the prediction of the SM usage and hence the allocated SM numbers respectively. The figure shows that \method\ provides adaptive SM schedule for the RAN in a complex scenario.
The per-slot level figure zooms into the sub-millisecond level and demonstrates that \method\ successfully predicts and schedules the SM budget for the RAN per slot.}

\subsection{Scheduling Unseen OTA and Multi-cell Scenarios}

\textbf{Unseen OTA AI-RAN Scenario:}We deploy \method\ on the private 5G network setup and demonstrate a single-cell AI-RAN case. The AI-RAN serves 4 OTA UEs in the cell and concurrently runs TimelyLLM. The UEs run fluctuated bi-directional traffic flows, and the TimelyLLM uses a 70B parameter Llama model.
This scenario is demonstrated in Fig.~\ref{fig:eval_fullstack}, the comparison between \method\ and no GPU SM management. Therefore we omit the description here for brevity.

\noindent
\textbf{Unseen Multi-cell AI-RAN Scenario:}
We further scale up the test cases to multiple cells accelerated by the GPU for the evaluation. 
We use TestMAC and RU emulator to build scaled-up multi-cell unseen scenarios for the testing, supporting up to 27 concurrent cells, considering the limitation on building OTA networks with such number of cells.
Fig.~\ref{fig:eval_multicell} shows a dynamic test case of the AI-RAN running \textcolor{black}{up to 27 cells and the TimelyLLM robotic application concurrently. There are up to 138 UEs with DL traffic and up to 61 UEs with UL traffic in this fluctuated scenario. The TimelyLLM runs the Llama-3.1-8B (AWQ-INT4) model.} We compare \method\ with MPS clients acquiring static SM partition according to their priorities on these two processes. 
\method\ applies the generalized model with the weights trained in \S\ref{sec:eval_train}. Correspondingly, we evaluate two types of SM partition for MPS clients, one prioritizing the RAN process, the other preferring TimelyLLM.
Fig.~\ref{fig:eval_multicell}(a) demonstrates the SM scheduling patterns of \method\ and static MPS client partition. \method\ schedules the SM budget adaptively, according to the prediction. 
The RAN prioritized partition assigns 108 SMs to the RAN and 24 SMs to TimelyLLM. Meanwhile, the AI prioritized partition assigns 52 SMs to the RAN and 80 SMs to TimelyLLM.
Fig.~\ref{fig:eval_multicell}(b) shows that \method\ and the static partition with RAN priority provide almost the same goodput for the UL process. In contrast, the static partition with AI priority noticeably jeopardizes the RAN UL goodput.
Fig.~\ref{fig:eval_multicell}(c) shows that the static MPS client partition prioritizing AI increases the UL latency significantly, which severely harms the RAN stability. 
Fig.~\ref{fig:eval_multicell}(d) shows the Time-to-First-Action (TTFA) comparison of TimelyLLM. The static MPS client partition with AI priority severely jeopardizes the RAN stability to reach low TTFA, which is not practical. Consequently, \method\ provides lower TTFA for TimelyLLM with protection on RAN latency, compared with the static partition with RAN priority.

\begin{table}[!h]
    \caption{\method~latency \textcolor{black}{overhead} evaluation under different cell numbers (\#Cells) and DL/UL throughput (Tput) scenarios, reported as (\textbf{\textit{mean $\pm$ std}}), showing negligible overhead and scalability.}
    \centering
    \footnotesize
    \setlength{\tabcolsep}{3pt}
    \begin{tabular*}{\columnwidth}{@{\extracolsep{\fill}}cccccc@{}}
    \toprule
     & \multicolumn{2}{c}{\textbf{Tput (Mbps)}} & \multicolumn{3}{c}{\textbf{Latency ($\mu$s)}} \\
    \cmidrule(lr){2-3} \cmidrule(l){4-6}
    \textbf{\#Cells} & \textbf{DL} & \textbf{UL}
     & \textbf{\begin{tabular}[b]{@{}c@{}}RAN\\Monitor\end{tabular}}
     & \textbf{\begin{tabular}[b]{@{}c@{}}SM\\Predictor\end{tabular}}
     & \textbf{\begin{tabular}[b]{@{}c@{}}SM\\Allocator\end{tabular}} \\
    \midrule
    8 & 5266$\pm$887 & 266$\pm$59 & 0.68$\pm$0.13 & 22.89$\pm$0.60 & 0.50$\pm$0.31 \\
    16 & 7284$\pm$2638 & 379$\pm$160 & 0.71$\pm$0.11 & 20.79$\pm$0.85 & 0.46$\pm$0.21 \\
    24 & 11228$\pm$4059 & 531$\pm$206 & 0.74$\pm$0.14 & 24.41$\pm$1.22 & 0.48$\pm$0.16 \\
    27 & 11483$\pm$4910 & 594$\pm$244 & 0.70$\pm$0.16 & 22.14$\pm$1.03 & 0.52$\pm$0.16 \\
    \bottomrule
    \end{tabular*}
    \label{tab:eval_lat}
\end{table}

\subsection{Latency Overhead}

We measure the latency overhead of \method~in the AI-RAN system at the run time, shown in Table~\ref{tab:eval_lat}. The latency ablation presents four scaled-up multi-cell cases, from 8 to 27 concurrent cells. Each case runs fluctuated bi-directional (DL/UL) traffic. The throughput fluctuations are due to the dynamic traffic patterns and varying active-cell count. Correspondingly, the \rma~and \saa~modules show the latency of lower than 1 $\mu$s. The latency of the \spa~module with \textcolor{black}{a four MLP ensemble} inference on CPU costs less than 25 $\mu$s. The total latency of \method\ is less than 27 $\mu$s in different multi-cell cases, demonstrating the negligible overhead with scalability.


\section{Conclusion and Future Works}
\label{sec:conc}
In this paper, we propose \method, an application-transparent scalable AI-RAN GPU SM scheduler that coordinates both RAN process and AI application dynamically in real time. \method\ characterizes the AI-RAN GPU SM usage, predicts the next-slot SM usage, and allocates the SM resource to both RAN and AI. We demonstrate the effectiveness of \method\ using OTA AI-RAN and multi-cell AI-RAN setup co-located with TimelyLLM. \method\ provides significant advantage over the static MPS client SM partition on AI-RAN management. However, it remains an early prototype. 
\method\ is a first step toward a general AI-RAN GPU sharing framework, which must support the full range of AI-RAN use cases. In AI-on-RAN, AI workloads serve RAN users and must meet their own quality-of-service and latency requirements, so the framework must transparently infer each application's requirements and arbitrate among multiple co-located models. In AI-for-RAN, AI
models directly drive RAN decisions in real time, either within the PHY pipeline or through disaggregated interfaces such as the real-time RIC; the framework must therefore allocate resources to these models at the same priority as RAN processing, ensuring RAN stability and performance. Our future work will extend \method\ to support these use cases.

\bibliographystyle{ACM-Reference-Format}
\bibliography{references} 

\end{document}